\font\twelverm=cmr12
\font\elfrm=cmr10 scaled 1097
\font\tenrm=cmr10
\font\tentt=cmtt10
\font\tenit=cmti10
\font\tensl=cmsl10
\font\elfsl=cmsl10 scaled 1097
\font\ninerm=cmr9
\font\eightrm=cmr8
\font\sixrm=cmr6
\font\twelvei=cmmi12
\font\twelvebi=cmmib10 scaled\magstep1
\font\tenbi=cmmib10
\font\elfi=cmmi10 scaled 1097
\font\ninei=cmmi9
\font\eighti=cmmi8
\font\sixi=cmmi6
\font\twelvesy=cmsy10 scaled\magstep1
\font\elfsy=cmsy10 scaled 1097
\font\ninesy=cmsy9
\font\eightsy=cmsy8
\font\sixsy=cmsy6
\font\twelvebf=cmbx12
\font\elfbf=cmbx10 scaled 1097
\font\ninebf=cmbx9
\font\eightbf=cmbx8
\font\sixbf=cmbx6
\font\twelvett=cmtt12
\font\ninett=cmtt9
\font\eighttt=cmtt8
\font\twelveit=cmti12
\font\elfit=cmti10 scaled 1097
\font\nineit=cmti9
\font\eightit=cmti8
\font\twelvesl=cmsl12
\font\ninesl=cmsl9
\font\eightsl=cmsl8
\font\twelveex=cmex10 scaled\magstep1
\font\bsi=cmbsy10 scaled\magstep1
\font\si=cmbsy10
\newskip\ttglue
\def\twelvepoint{\def\rm{\fam0\twelverm}%
\def\sl{\fam\slfam\twelvesl}%
\textfont0=\twelverm   \scriptfont0=\ninerm  \scriptscriptfont0=\sevenrm%
\textfont1=\twelvei    \scriptfont1=\ninei   \scriptscriptfont1=\seveni%
\textfont2=\twelvesy   \scriptfont2=\ninesy  \scriptscriptfont2=\sevensy%
\textfont3=\twelveex   \scriptfont3=\twelveex\scriptscriptfont3=\twelveex%
\textfont\itfam=\twelveit    \def\it{\fam\itfam\twelveit}%
\textfont\slfam=\twelvesl    \def\sl{\fam\slfam\twelvesl}%
\textfont\ttfam=\twelvett    \def\tt{\fam\ttfam\twelvett}%
\textfont\bffam=\twelvebf   \scriptfont\bffam=\ninebf%
\scriptscriptfont\bffam=\sevenbf     \def\bf{\fam\bffam\twelvebf}%
\ttglue=.5em plus .25em minus.15em%
\normalbaselineskip=15pt\rm%
\setbox\strutbox=\hbox{\vrule height9.75pt depth4pt width0pt}}
\def\elfpoint{\def\rm{\fam0\elfrm}%
\def\sl{\fam\slfam\elfsl}%
\textfont0=\elfrm\scriptfont0=\ninerm\scriptscriptfont0=\sixrm%
\textfont1=\elfi\scriptfont1=\ninei\scriptscriptfont1=\sixi%
\textfont2=\elfsy\scriptfont2=\ninesy\scriptscriptfont2=\sixsy%
\textfont3=\tenex\scriptfont3=\tenex\scriptscriptfont3=\tenex%
\textfont\itfam=\elfit\def\it{\fam\itfam\elfit}%
\scriptscriptfont\bffam=\sixbf\def\bf{\fam\bffam\elfbf}%
\ttglue=.45em plus.25em minus.15em%
\normalbaselineskip=13.5pt\rm%
\setbox\strutbox=\hbox{\vrule height7.8pt depth3.15pt width0pt}}

\def\tenpoint{\def\rm{\fam0\tenrm}%
\def\sl{\fam\slfam\tensl}%
\def\tt{\fam\ttfam\tentt}%
\textfont0=\tenrm\scriptfont0=\sevenrm\scriptscriptfont0=\fiverm%
\textfont1=\teni\scriptfont1=\seveni\scriptscriptfont1=\fivei%
\textfont2=\tensy\scriptfont2=\sevensy\scriptscriptfont2=\fivesy%
\textfont3=\tenex\scriptfont3=\tenex\scriptscriptfont3=\tenex%
\textfont\itfam=\tenit\def\it{\fam\itfam\tenit}%
\textfont\ttfam=\tentt\def\tt{\fam\ttfam\tentt}%
\scriptscriptfont\bffam=\fivebf\def\bf{\fam\bffam\tenbf}%
\ttglue=.5em plus.25em minus.15em%
\normalbaselineskip=12.5pt\rm%
\setbox\strutbox=\hbox{\vrule height8.5pt depth3.5pt width0pt}}

\def\ninepoint{\def\rm{\fam0\ninerm}
 \textfont0=\ninerm   \scriptfont0=\sixrm \scriptscriptfont0=\fiverm
 \textfont1=\ninei    \scriptfont1=\sixi  \scriptscriptfont1=\fivei
 \textfont2=\ninesy   \scriptfont2=\sixsy \scriptscriptfont2=\fivesy
 \textfont3=\tenex    \scriptfont3=\tenex \scriptscriptfont3=\tenex
 \textfont\itfam=\nineit    \def\it{\fam\itfam\nineit}
 \textfont\slfam=\ninesl    \def\sl{\fam\slfam\ninesl}
 \textfont\ttfam=\ninett    \def\tt{\fam\ttfam\ninett}
 \textfont\bffam=\ninebf   \scriptfont\bffam=\sixbf
  \scriptscriptfont\bffam=\fivebf     \def\bf{\fam\bffam\ninebf}
\ttglue=.5em plus .25em minus.15em
\normalbaselineskip=11pt\rm
\setbox\strutbox=\hbox{\vrule height7.7pt depth3pt width0pt}}

\def\eightpoint{\def\rm{\fam0\eightrm}
 \textfont0=\eightrm  \scriptfont0=\sixrm  \scriptscriptfont0=\fiverm
 \textfont1=\eighti   \scriptfont1=\sixi   \scriptscriptfont1=\fivei
 \textfont2=\eightsy  \scriptfont2=\sixsy  \scriptscriptfont2=\fivesy
 \textfont3=\tenex    \scriptfont3=\tenex  \scriptscriptfont3=\tenex
 \textfont\itfam=\eightit  \def\it{\fam\itfam\eightit}
 \textfont\slfam=\eightsl  \def\sl{\fam\slfam\eightsl}
 \textfont\ttfam=\eighttt  \def\tt{\fam\ttfam\eighttt}
 \textfont\bffam=\eightbf  \scriptfont\bffam=\sixbf
  \scriptscriptfont\bffam=\fivebf  \def\bf{\fam\bffam\eightbf}
\ttglue=.5em plus.25em minus.15em
\normalbaselineskip=9pt\rm
\setbox\strutbox=\hbox{\vrule height7pt depth2pt width0pt}}

\font\rmb=cmr12 scaled \magstep 1
 1
\font\rmc=cmr12 scaled \magstep 2
\font\rmd=cmr12 scaled \magstep 3
\font\rme=cmr12 scaled \magstep 4
 2
\font\ita=cmti12
\font\itb=cmti12 scaled \magstep 1
 3
 4
 5
\font\mitb=cmmi10 scaled \magstep 2
\font\mitc=cmmi10 scaled \magstep 3
\font\mitd=cmmi10 scaled \magstep 4
 5

 1
\def\title#1#2#3{\par\ifnum\prevgraf>0 \npageno=0 \fi
   \global\shortno=0\global\secno=0\global\tabno=0\global
   \figno=0\global\footno=0\global\refno=0\goodbreak\ifnum\npageno=0
   \par\vskip 2cm plus2.5mm minus2.5mm\fi{\noindent\textfont1=\mitd
   \rmd #1$\vphantom{\hbox{\rmd y}}$}\par
   \penalty1000\vskip8pt\noindent{\itb #2$\vphantom{\hbox{\itb y}}$}{\par
   \penalty1000\vskip7.5pt\baselineskip11.5pt\hrule height 0.1pt\par
   \penalty1000\vskip7pt{\tenpoint\noindent\rm #3$\vphantom{y}$}\par
   \penalty1000\vskip6pt\hrule height 0.1pt
   \par\penalty1000\vskip9pt\noindent}}
\def\stitle#1#2#3{\par\ifnum\prevgraf>0 \npageno=0 \fi
   \global\shortno=0\global\secno=0\global\tabno=0\global
   \figno=0\global\footno=0\global\refno=0\goodbreak\ifnum\npageno=0
   \par\vskip 1.6cm plus2mm minus2mm\fi{\noindent\textfont1=\mitc
   \rmc #1$\vphantom{\hbox{\rmc y}}$}\par
   \penalty1000\vskip6.5pt\noindent{\ita #2$\vphantom{\hbox{\itb y}}$}{\par
   \penalty1000\vskip6pt\baselineskip10pt\hrule height 0.1pt\par
   \penalty1000\vskip5.75pt{\ninepoint\noindent\rm #3$\vphantom{y}$}\par
   \penalty1000\vskip4.8pt\hrule height 0.1pt
   \par\penalty1000\vskip7.25pt\noindent}}
\def\supertitle#1#2#3#4{\par\ifnum\prevgraf>0 \npageno=0 \fi
   \global\shortno=0\global\secno=0\global\tabno=0\global
   \figno=0\global\footno=0\global\refno=0\goodbreak\ifnum\npageno=0
   \par\vskip 1.3cm plus2.5mm minus2.5mm\fi\noindent{\textfont1=\mitc
   \rmc #1}\par
   \penalty1000\vskip1cm plus 3mm minus 7mm{\noindent\textfont1=\mitd\rmd
   #2$\vphantom{\hbox{\rmd y}}$}\par\penalty1000\vskip8pt\noindent{\itb
   #3$\vphantom{\hbox{\itb y}}$}{\par\penalty1000\vskip7.5pt\baselineskip
   11.5pt\hrule height 0.1pt\par\penalty1000\vskip7pt{\tenpoint\noindent\rm
   #4$\vphantom{y}$}\par\penalty1000\vskip6pt\hrule height 0.1pt\par
   \penalty1000\vskip9pt\noindent}}
\def\ssupertitle#1#2#3#4{\par\ifnum\prevgraf>0 \npageno=0 \fi
   \global\shortno=0\global\secno=0\global\tabno=0\global
   \figno=0\global\footno=0\global\refno=0\goodbreak\ifnum\npageno=0
   \par\vskip 1.1cm plus2.1mm minus2.1mm\fi\noindent{\textfont1=\mitb
   \rmb #1$\vphantom{\hbox{\rmb y}}$}\par
   \penalty1000\vskip8mm plus 3mm minus 6mm{\noindent\textfont1=\mitc\rmc
   #2$\vphantom{\hbox{\rmc y}}$}\par\penalty1000\vskip8pt\noindent{\ita
   #3$\vphantom{\hbox{\ita y}}$}{\par\penalty1000\vskip6pt\baselineskip
   10pt\hrule height
   0.1pt\par\penalty1000\vskip5.5pt{\ninepoint\noindent\rm
   #4$\vphantom{y}$}\par\penalty1000\vskip6pt\hrule height 0.1pt\par
   \penalty1000\vskip9pt\noindent}}
%
%
\def\firstreftitle#1#2#3#4{\par\ifnum\prevgraf>0 \npageno=0 \fi
   \global\shortno=0\global\secno=0\global\tabno=0\global
   \figno=0\global\footno=0\global\refno=0\goodbreak\ifnum\npageno=0
   \par\vskip 1.5cm plus2.5mm minus2.5mm\fi\noindent{\textfont1=\mitd
   \rmc Progress in Meteor Science}{\par\penalty1000
\vskip2mm plus 0.5mm minus 0.5mm\baselineskip11.5pt{\tenpoint\noindent\it
Articles in this section have been formally refereed by at least one
professional and one experienced, knowledgeable amateur meteor worker,
and deal with global analyses of meteor data, methods for meteor
observing and data reduction, observations with professional equipment, or
theoretical studies.}\par}\par\penalty1000
\vskip7.5mm plus 2.5mm minus 2.5mm{\noindent\textfont1=\mitd\rmd
#1$\vphantom{\hbox{\rmd y}}$}\par\penalty1000\vskip8pt\noindent{\itb
#2$\vphantom{\hbox{\itb y}}$}{\tenpoint\rm
\baselineskip11.5pt\footnote{}{\hskip-\parindent #3\smallnewpar\it
WGN, the Journal of the
International Meteor Organization, Vol.~\the\volno, No.~\the\nrno,
\month~\the\yearno, pp.~\the\firstpageno--\the\lastpageno.}}%
{\par\penalty1000\vskip7.5pt\baselineskip
11.5pt\hrule height 0.1pt\par\penalty1000\vskip7pt{\tenpoint\noindent\rm
#4$\vphantom{y}$}\par\penalty1000\vskip6pt\hrule height 0.1pt\par
\penalty1000\vskip9pt\noindent}}%
\def\reftitle#1#2#3#4{\par\ifnum\prevgraf>0 \npageno=0 \fi
   \global\shortno=0\global\secno=0\global\tabno=0\global
   \figno=0\global\footno=0\global\refno=0\goodbreak\ifnum\npageno=0
   \par\vskip 1.5cm plus2.5mm minus2.5mm\fi\noindent{\textfont1=\mitd\rmd
#1$\vphantom{\hbox{\rmd y}}$}\par\penalty1000\vskip8pt\noindent{\itb
#2$\vphantom{\hbox{\itb y}}$}{\tenpoint\rm
\baselineskip11.5pt\footnote{}{\hskip-\parindent #3\smallnewpar\it
WGN, the Journal of the
International Meteor Organization, Vol.~\the\volno, No.~\the\nrno,
\month~\the\yearno, pp.~\the\firstpageno--\the\lastpageno.}}%
{\par\penalty1000\vskip7.5pt\baselineskip
11.5pt\hrule height 0.1pt\par\penalty1000\vskip7pt{\tenpoint\noindent\rm
#4$\vphantom{y}$}\par\penalty1000\vskip6pt\hrule height 0.1pt\par
\penalty1000\vskip9pt\noindent}}%
\def\noabsupertitle#1#2#3{\par\ifnum\prevgraf>0 \npageno=0 \fi
   \global\shortno=0\global\secno=0\global\tabno=0\global
   \figno=0\global\footno=0\global\refno=0\goodbreak\ifnum\npageno=0
   \par\vskip 2cm plus2.5mm minus2.5mm\fi\noindent{\rmc #1}\par
   \penalty1000\vskip1cm plus 5mm minus 5mm{\noindent\rmd
   #2$\vphantom{\hbox{\rmd y}}$}\par\penalty1000\vskip8pt\noindent{\itb
   #3$\vphantom{\hbox{\itb y}}$}{\par\penalty1000\vskip7.5pt
   \hrule height 0.1pt\par\penalty1000\vskip9pt\noindent}}
\def\noabtitle#1#2{\par\ifnum\prevgraf>0 \npageno=0\fi
   \global\shortno=0\global\secno=0\global\tabno=0\global
   \figno=0\global\footno=0\global\refno=0\goodbreak\ifnum\npageno=0
   \par\vskip2cm plus 2.5mm minus 2.5mm\fi
   {\noindent\rmd #1$\vphantom{\hbox{\rme y}}$}\par\penalty1000\vskip8pt
   \noindent{\itb #2$\vphantom{\hbox{\itb y}}$}\par\penalty1000\vskip7.5pt
   \hrule height0.1pt\par\penalty1000\vskip9pt\noindent}
\def\snoabtitle#1#2{\par\ifnum\prevgraf>0 \npageno=0\fi
   \global\shortno=0\global\secno=0\global\tabno=0\global
   \figno=0\global\footno=0\global\refno=0\goodbreak\ifnum\npageno=0
   \par\vskip1.6cm plus 2mm minus 2mm\fi
   {\noindent\textfont1=\mitc
   \rmc #1$\vphantom{\hbox{\rmc y}}$}\par\penalty1000\vskip6.5pt
   \noindent{\ita #2$\vphantom{\hbox{\ita y}}$}\par\penalty1000\vskip6pt
   \hrule height0.1pt\par\penalty1000\vskip7.25pt\noindent}
\def\shorttitle#1#2{\global\secno=0\global\tabno=0\global
   \figno=0\global\footno=0\global\refno=0{\noindent
   \rmd #1$\vphantom{\hbox{\rme y}}$}\par
   \penalty1000\vskip8pt\noindent{\itb #2$\vphantom{\hbox{\itb y}}$}\par
   \penalty1000\vskip7.5pt\hrule height0.1pt\par\penalty1000
   \vskip9pt\noindent}
\def\sshorttitle#1#2{\global\secno=0\global\tabno=0\global
   \figno=0\global\footno=0\global\refno=0{\noindent
   \rmc #1$\vphantom{\hbox{\rmd y}}$}\par
   \penalty1000\vskip6.5pt\noindent{\ita #2$\vphantom{\hbox{\ita y}}$}\par
   \penalty1000\vskip6pt\hrule height0.1pt\par\penalty1000
   \vskip7.25pt\noindent}
\def\noauttitle#1{\global\secno=0\global\tabno=0\global\figno=0
 \global\footno=0\global\refno=0{\noindent\rmd #1$\vphantom{\hbox{\rmd y}}$}
   \par\vskip 7.5pt\hrule height
   0.1pt\par\penalty1000\vskip9pt\noindent}
\def\noabsemibigtitle#1#2#3#4{\global\shortno=0\global\secno=0
   \global\tabno=0\global\figno=0\global\footno=0\global\refno=0\goodbreak
   \ifnum\npageno=0 \par\vskip 2cm plus2.5mm  minus2.5mm\fi{\noindent
   \rmc #1$\vphantom{\hbox{\rmd y}}$}\par
   \penalty1000\vskip7pt\noindent{\rmd #2$\vphantom{\hbox{\rmc y}}$}\par
   \penalty1000\vskip7pt\noindent{\rmd #3$\vphantom{\hbox{\rmc y}}$}\par
 \par\penalty1000\vskip7pt\noindent{\itb #4$\vphantom{\hbox{\itb y}}$}{\par
   \penalty1000\vskip7.5pt\baselineskip11.5pt\hrule height 0.1pt\par
   \penalty1000\vskip9pt\noindent}}
\def\supersemititle#1#2#3#4#5{\global\secno=0\global\tabno=0\global\figno=0
   \global\footno=0\global\refno=0\goodbreak\ifnum\npageno=0
   \par\vskip 1.3cm plus2.5mm minus2.5mm\fi\noindent{\textfont1=\mitd
   \rmc #1$\vphantom{\hbox{\rmc y}}$}\par
   \penalty1000\vskip7pt\noindent{\textfont1=\mitd\rmd
   #2$\vphantom{\hbox{\rmd y}}$}\par
   \penalty1000\vskip7pt\noindent{\rmd #3$\vphantom{\hbox{\rmd y}}$}\par
 \par\penalty1000\vskip8pt\noindent{\itb #4$\vphantom{\hbox{\itb y}}$}{\par
   \penalty1000\vskip7.5pt\baselineskip11.5pt\hrule height 0.1pt\par
   \penalty1000\vskip7pt{\tenpoint\noindent\rm
   #5$\vphantom{y}$}\par\penalty1000\vskip6pt\hrule height 0.1pt\par
   \penalty1000\vskip9pt\noindent}}
\def\noabsupersemititle#1#2#3#4{\global\secno=0\global\tabno=0\global\figno=0
   \global\footno=0\global\refno=0\goodbreak\ifnum\npageno=0
   \par\vskip 1.3cm plus2.5mm minus2.5mm\fi\noindent{\textfont1=\mitd
   \rmc #1$\vphantom{\hbox{\rmc y}}$}\par
   \penalty1000\vskip7pt\noindent{\textfont1=\mitd\rmd
   #2$\vphantom{\hbox{\rmd y}}$}\par
   \penalty1000\vskip7pt\noindent{\rmd #3$\vphantom{\hbox{\rmd y}}$}\par
 \par\penalty1000\vskip8pt\noindent{\itb #4$\vphantom{\hbox{\itb y}}$}{\par
   \penalty1000\vskip7.5pt\baselineskip11.5pt\hrule height 0.1pt\par
   \penalty1000\vskip9pt\noindent}}
\def\ssupersemititle#1#2#3#4#5{\global\secno=0\global\tabno=0\global\figno=0
   \global\footno=0\global\refno=0\goodbreak\ifnum\npageno=0
   \par\vskip 1.3cm plus2.5mm minus2.5mm\fi\noindent{\textfont1=\mitb
   \rmb #1$\vphantom{\hbox{\rmb y}}$}\par
   \penalty1000\vskip6.5pt\noindent{\rmc #2$\vphantom{\hbox{\rmc y}}$}\par
   \penalty1000\vskip6.5pt\noindent{\rmb #3$\vphantom{\hbox{\rmb y}}$}\par
 \par\penalty1000\vskip8pt\noindent{\ita #4$\vphantom{\hbox{\ita y}}$}{\par
   \penalty1000\vskip7pt\baselineskip10.35pt\hrule height 0.1pt\par
   \penalty1000\vskip6pt{\ninepoint\noindent\rm
   #5$\vphantom{y}$}\par\penalty1000\vskip5.75pt\hrule height 0.1pt\par
   \penalty1000\vskip8pt\noindent}}

\def\bigtitle#1#2#3#4{\par\ifnum\prevgraf>0 \npageno=0 \fi
   \global\shortno=0\global\secno=0\global\tabno=0\global
   \figno=0\global\footno=0\global\refno=0\goodbreak\ifnum\npageno=0
   \par\vskip 2cm plus2.5mm  minus2.5mm\fi{\noindent\textfont1=\mitd\rmd #1$
   \vphantom{\hbox{\rmd y}}$}\par\penalty1000\vskip8pt\noindent
   {\textfont1=\mitd\rmd #2$\vphantom{\hbox{\rmd y}}$}
\par\penalty1000\vskip 8pt\noindent{\itb #3$\vphantom{\hbox{\itb y}}$}{\par
   \penalty1000\vskip7.5pt\baselineskip11.5pt\hrule height 0.1pt\par
   \penalty1000\vskip 7pt{\tenpoint\noindent\rm #4$\vphantom{y}$}\par
   \penalty1000\vskip6pt\hrule height 0.1pt\par\vskip9pt\noindent}}
\def\noabbigtitle#1#2#3{\global\shortno=0\global\secno=0\global\tabno=0\global
   \figno=0\global\footno=0\global\refno=0\goodbreak\ifnum\npageno=0
   \par\vskip 2cm plus2.5mm  minus2.5mm\fi{\noindent
   \rmd #1$\vphantom{\hbox{\rmd y}}$}\par
   \penalty1000\vskip8pt\noindent{\rmd #2$\vphantom{\hbox{\rmd y}}$}
 \par\penalty1000\vskip 8pt\noindent{\itb #3$\vphantom{\hbox{\itb y}}$}{\par
   \penalty1000\vskip7.5pt\hrule height 0.1pt\par
   \penalty1000\vskip9pt\noindent}}
\def\snoabbigtitle#1#2#3{\global\shortno=0
   \global\secno=0\global\tabno=0\global
   \figno=0\global\footno=0\global\refno=0\goodbreak\ifnum\npageno=0
   \par\vskip 1.6cm plus2mm  minus2mm\fi{\noindent\textfont1=\mitc%
   \rmc #1$\vphantom{\hbox{\rmc y}}$}\par
   \penalty1000\vskip6pt\noindent{\rmc #2$\vphantom{\hbox{\rmc y}}$}
 \par\penalty1000\vskip 6pt\noindent{\ita #3$\vphantom{\hbox{\ita y}}$}{\par
   \penalty1000\vskip6pt\hrule height 0.1pt\par
   \penalty1000\vskip7.25pt\noindent}}
\def\snoabsemititle#1#2#3{\global\shortno=0
   \global\secno=0\global\tabno=0\global
   \figno=0\global\footno=0\global\refno=0\goodbreak\ifnum\npageno=0
   \par\vskip 1.6cm plus2mm  minus2mm\fi{\noindent\textfont1=\mitc%
   \rmc #1$\vphantom{\hbox{\rmc y}}$}\par
   \penalty1000\vskip6pt\noindent{\rmb #2$\vphantom{\hbox{\rmb y}}$}
   \par\penalty1000\vskip 6pt\noindent{\ita #3$\vphantom{\hbox{\ita y}}$}{\par
   \penalty1000\vskip6pt\hrule height 0.1pt\par
   \penalty1000\vskip7.25pt\noindent}}
\def\snoabsupertitle#1#2#3{\global\shortno=0
   \global\secno=0\global\tabno=0\global
   \figno=0\global\footno=0\global\refno=0\goodbreak\ifnum\npageno=0
   \par\vskip 1.6cm plus2mm  minus2mm\fi{\noindent\textfont1=\mitb%
   \rmb #1$\vphantom{\hbox{\rmb y}}$}\par
   \penalty1000\vskip6pt\noindent{\rmc #2$\vphantom{\hbox{\rmc y}}$}
   \par\penalty1000\vskip 6pt\noindent{\ita #3$\vphantom{\hbox{\ita y}}$}{\par
   \penalty1000\vskip6pt\hrule height 0.1pt\par
   \penalty1000\vskip7.25pt\noindent}}
\def\snoabbigsemititle#1#2#3#4{\global\shortno=0
   \global\secno=0\global\tabno=0\global
   \figno=0\global\footno=0\global\refno=0\goodbreak\ifnum\npageno=0
   \par\vskip 1.6cm plus2mm  minus2mm\fi{\noindent
   \rmc #1$\vphantom{\hbox{\rmc y}}$}\par
   \penalty1000\vskip6pt{\noindent
   \rmc #2$\vphantom{\hbox{\rmc y}}$}\par
   \penalty1000\vskip6pt\noindent{\rmb #3$\vphantom{\hbox{\rmb y}}$}
 \par\penalty1000\vskip 6pt\noindent{\ita #4$\vphantom{\hbox{\ita y}}$}{\par
   \penalty1000\vskip6pt\hrule height 0.1pt\par
   \penalty1000\vskip7.25pt\noindent}}
\def\semititle#1#2#3#4{\global\shortno=0\global\secno=0\global\tabno=0\global
   \figno=0\global\footno=0\global\refno=0\goodbreak\ifnum\npageno=0
   \par\vskip 2cm plus2.5mm  minus2.5mm\fi{\noindent
   \rmd #1$\vphantom{\hbox{\rmd y}}$}\par
   \penalty1000\vskip7pt\noindent{\rmc #2$\vphantom{\hbox{\rmc y}}$}
 \par\penalty1000\vskip 7pt\noindent{\itb #3$\vphantom{\hbox{\itb y}}$}{\par
   \penalty1000\vskip7.5pt\baselineskip11.5pt\hrule height 0.1pt\par
   \penalty1000\vskip 7pt{\tenpoint\noindent\rm #4$\vphantom{y}$}\par
   \penalty1000\vskip6pt\hrule height 0.1pt\par\vskip9pt\noindent}}
\def\noabsemititle#1#2#3{\global\shortno=0\global\secno=0\global\tabno=0
    \global\figno=0\global\footno=0\global\refno=0\goodbreak
    \ifnum\npageno=0 \par\vskip 2cm plus2.5mm  minus2.5mm\fi{\noindent
    \rmd #1$\vphantom{\hbox{\rmd y}}$}\par
   \penalty1000\vskip7pt\noindent{\rmc #2$\vphantom{\hbox{\rmc y}}$}
 \par\penalty1000\vskip7pt\noindent{\itb #3$\vphantom{\hbox{\itb y}}$}{\par
   \penalty1000\vskip7.5pt\baselineskip11.5pt\hrule height 0.1pt\par
   \penalty1000\vskip9pt\noindent}}
\def\noautshortnote#1{\global\secno=0\global\tabno=0\global
   \figno=0\global\footno=0\global\refno=0\goodbreak\ifnum\npageno=0
   \ifnum\shortno=0 \vskip6mm plus 2mm minus 2mm\noindent
   \hrule height 1.2pt\par\vskip8.5mm plus 2.5mm minus 2.5mm
   \else\vskip17mm plus 4mm minus 3mm\fi\fi
   \global\shortno=1\noindent\noauttitle{#1}}
\def\shortnote#1#2{\global\secno=0\global\tabno=0\global
   \figno=0\global\footno=0\global\refno=0\goodbreak\ifnum\npageno=0
   \ifnum\shortno=0 \vskip6mm plus 2mm minus 2mm\noindent
   \hrule height 1.2pt\par\vskip8.5mm plus 2.5mm minus 2.5mm
   \else\vskip17mm plus 4mm minus 3mm\fi\fi
   \global\shortno=1\noindent\shorttitle{#1}{#2}}
\def\sshortnote#1#2{\global\secno=0\global\tabno=0\global
   \figno=0\global\footno=0\global\refno=0\goodbreak\ifnum\npageno=0
   \ifnum\shortno=0 \vskip5mm plus 1.75mm minus 1.75mm\noindent
   \hrule height 1.2pt\par\vskip7.25mm plus 2mm minus 2mm
   \else\vskip15.5mm plus 3.5mm minus 2.5mm\fi\fi
   \global\shortno=1\noindent\sshorttitle{#1}{#2}}
\def\shortnotes#1#2{\global\secno=0\global\tabno=0\global
   \figno=0\global\footno=0\global\refno=0\goodbreak\ifnum\npageno=0
   \ifnum\shortno=0 \vskip8.5mm plus 2.5mm minus 2.5mm
   \noindent\hrule height 1.2pt\par\vskip8.5mm plus 2.5mm minus 2.5mm
   \else\vskip17mm plus 5mm minus 5mm\fi\fi
 \global\shortno=1\noindent{\rmc Short Notes}\par\penalty10000\vskip8.5mm
    plus 2.5mm minus 2.5mm\noindent\shorttitle{#1}{#2}}
\parskip=0pt
\parindent=25pt
\def\newpar{\edef\next{\hangafter=\the\hangafter
    \hangindent=\the\hangindent}
    \par\penalty10000\ifnum\prevgraf>0 \global\npageno=0\fi
    \par\penalty-100\vskip 4pt plus 2pt minus 2pt\next%
    \edef\next{\ifnum\prevgraf>-\the\hangafter\prevgraf=0\hangafter=1
         \hangindent=0pt\else\prevgraf=\the\prevgraf\fi}
    \noindent\next}
\def\smallnewpar{\edef\next{\hangafter=\the\hangafter
    \hangindent=\the\hangindent}
    \par\penalty10000\ifnum\prevgraf>0 \global\npageno=0\fi
    \par\penalty-100\vskip 3pt plus 1.5pt minus 1pt\next%
    \edef\next{\ifnum\prevgraf>-\the\hangafter\prevgraf=0\hangafter=1
         \hangindent=0pt\else\prevgraf=\the\prevgraf\fi}
    \noindent\next}
\def\pensmallnewpar{\edef\next{\hangafter=\the\hangafter
    \hangindent=\the\hangindent}
    \par\penalty10000\ifnum\prevgraf>0 \global\npageno=0\fi
    \par\penalty10000\vskip 3pt plus 1.5pt minus 1pt\next%
    \edef\next{\ifnum\prevgraf>-\the\hangafter\prevgraf=0\hangafter=1
         \hangindent=0pt\else\prevgraf=\the\prevgraf\fi}
    \noindent\next}
\newcount\npageno
\newcount\shortno
\newcount\secno
\newcount\tabno
\newcount\figno
\newcount\refno
\newdimen\figindent
\newdimen\figheight
\newdimen\newfigheight
\newdimen\figwidth
\npageno=0
\shortno=0
\secno=0
\refno=0
\tabno=0
\figno=0
\figindent=0sp
\figheight=0sp
\newfigheight=0sp
\figwidth=0sp
\def\section#1{\par\global\npageno=0
    \global\advance\secno by 1\ifnum\prevgraf=0 \relax
    \else\penalty-100\vskip 12pt plus 6pt minus 4.5pt\fi
    {\textfont0=\twelvebf\textfont1=\twelvebi\textfont2=\bsi
     \scriptfont0=\ninebf
    \noindent{\bf\the\secno.\ #1}}\par\penalty10000
    \vskip 4pt plus 2pt minus2pt\noindent}%
\def\smallsection#1{\par\global\npageno=0
    \global\advance\secno by 1\ifnum\prevgraf=0 \relax
    \else\penalty-100\vskip 10pt plus 5pt minus 4pt\fi
    {\textfont0=\tenbf\textfont1=\tenbi\textfont2=\si
    \noindent{\bf\the\secno.\ #1}}\newpar}
\def\appsection#1{\par\global\npageno=0
   \ifnum\prevgraf=0 \relax
   \else\penalty-100\vskip 12pt plus 6pt minus 4.5pt\fi
   {\textfont0=\twelvebf\textfont1=\twelvebi\textfont2=\bsi
   \noindent{\bf#1}}\newpar}
\def\smallappsection#1{\par\global\npageno=0
   \ifnum\prevgraf=0 \relax
   \else\penalty-100\vskip 10pt plus 5pt minus 4pt\fi
   {\textfont0=\tenbf\textfont1=\tenbi\textfont2=\si
   \noindent{\bf#1}}\newpar}
\def\artref#1#2#3#4{\par\global\advance\refno by 1\ifnum\prevgraf=0 \relax
    \else\penalty-100\vskip 2pt plus 0.5pt minus 0.5pt\fi
    \hangindent\parindent\noindent\hbox to \parindent{[$
    \the\refno$]\hfill}#1, ``#2'', {\it #3\/}, #4.}
\def\sartref#1#2#3#4{\par\global\advance\refno by 1\ifnum\prevgraf=0 \relax
    \else\penalty-100\vskip 1.5pt plus 0.5pt minus 0.5pt\fi
    \hangindent\parindent\noindent\hbox to \parindent{[$
    \the\refno$]\hfill}#1, ``#2'', {\it #3\/}, #4.}
\def\noautref#1#2#3{\par\global\advance\refno by 1\ifnum\prevgraf=0 \relax
    \else\penalty-100\vskip 2pt plus 0.5pt minus 0.5pt\fi
    \hangindent\parindent\noindent\hbox to \parindent{[$
    \the\refno$]\hfill}``#1'', {\it #2\/}, #3.}
\def\noartref#1#2#3{\par\global\advance\refno by 1\ifnum\prevgraf=0 \relax
    \else\penalty-100\vskip 2pt plus 0.5pt minus 0.5pt\fi
    \hangindent\parindent\noindent\hbox to \parindent{[$
    \the\refno$]\hfill}#1, {\it #2\/}, #3.}
\def\snoartref#1#2#3{\par\global\advance\refno by 1\ifnum\prevgraf=0 \relax
    \else\penalty-100\vskip 1.5pt plus 0.5pt minus 0.5pt\fi
    \hangindent\parindent\noindent\hbox to \parindent{[$
    \the\refno$]\hfill}#1, {\it #2\/}, #3.}
\def\noartnoautref#1#2{\par\global\advance\refno by 1\ifnum\prevgraf=0 \relax
    \else\penalty-100\vskip 2pt plus 0.5pt minus 0.5pt\fi
    \hangindent\parindent\noindent\hbox to \parindent{[$
    \the\refno$]\hfill}{\it #1\/}, #2.}
\def\bookref#1#2#3{\par\global\advance\refno by 1\ifnum\prevgraf=0 \relax
    \else\penalty-100\vskip 2pt plus 0.5pt minus 0.5pt\fi
    \hangindent\parindent\noindent\hbox to \parindent{[$
    \the\refno$]\hfill}#1, ``#2'', #3.}
\def\sbookref#1#2#3{\par\global\advance\refno by 1\ifnum\prevgraf=0 \relax
    \else\penalty-100\vskip 1.5pt plus 0.5pt minus 0.5pt\fi
    \hangindent\parindent\noindent\hbox to \parindent{[$
    \the\refno$]\hfill}#1, ``#2'', #3.}
\def\noautbookref#1#2{\par\global\advance\refno by 1\ifnum\prevgraf=0 \relax
    \else\penalty-100\vskip 2pt plus 0.5pt minus 0.5pt\fi
    \hangindent\parindent\noindent\hbox to \parindent{[$
    \the\refno$]\hfill}``#1'', #2.}
\def\shortref#1#2{\par\global\advance\refno by 1\ifnum\prevgraf=0 \relax
    \else\penalty-100\vskip 2pt plus 0.5pt minus 0.5pt\fi
    \hangindent\parindent\noindent\hbox to \parindent{[$
    \the\refno$]\hfill}#1, ``#2''.}
\def\sourceref#1{\par\global\advance\refno by 1\ifnum\prevgraf=0 \relax
    \else\penalty-100\vskip 2pt plus 0.5pt minus 0.5pt\fi
    \hangindent\parindent\noindent\hbox to \parindent{[$
    \the\refno$]\hfill}{\it #1.}}
\def\perscom#1#2{\par\global\advance\refno by 1\ifnum\prevgraf=0 \relax
    \else\penalty-100\vskip 2pt plus 0.5pt minus 0.5pt\fi
    \hangindent\parindent\noindent\hbox to \parindent{[$
    \the\refno$]\hfill}#1, {\it personal communications\/}, #2.}
\def\libref#1#2{\par\hangindent\parindent\noindent
   \hbox to \parindent{$\bullet$\hfill}{\it #1}\pensmallnewpar{\rm #2}\newpar}

\newbox\tabbox
\def\table#1#2{\global\npageno=0\advance\tabno by 1
    \setbox\tabbox=\vbox{\tenpoint\tabskip=0pt
    \offinterlineskip\halign{#2}}
       $$\vbox{\tenpoint\baselineskip11.5pt\ialign{\hfill$##$\crcr
                  \vtop{\hsize=\wd\tabbox\hangindent1.65truecm\noindent%
                  \hbox to 1.65truecm{Table \the\tabno\ -- \hfill}#1}\crcr
                  \noalign{\vskip8pt}
                  \box\tabbox\crcr}}$$}
\def\nocaptable#1{\global\npageno=0\advance\tabno by 1
    \setbox\tabbox=\vbox{\tenpoint\tabskip=0pt
    \offinterlineskip\halign{#1}}
       $$\vbox{\tenpoint\baselineskip11.5pt\ialign{\hfill$##$\crcr
                  \box\tabbox\crcr}}$$
}\def\smalltable#1#2{\global\npageno=0\advance\tabno by 1
    \setbox\tabbox=\vbox{\ninepoint\tabskip=0pt
    \offinterlineskip\halign{#2}}
       $$\vbox{\ninepoint\baselineskip10.35pt\ialign{\hfill$##$\crcr
                  \vtop{\hsize=\wd\tabbox\hangindent1.5truecm\noindent%
      \hbox to 1.5truecm{\ninerm Table \the\tabno\ -- \hfill}#1}\crcr
                  \noalign{\vskip8pt}
                  \box\tabbox\crcr}}$$}
\def\nocapsmalltable#1{\global\npageno=0\advance\tabno by 1
    \setbox\tabbox=\vbox{\ninepoint\tabskip=0pt
    \offinterlineskip\halign{#1}}
       $$\vbox{\ninepoint\baselineskip10.35pt\ialign{\hfill$##$\crcr
                  \box\tabbox\crcr}}$$}
\def\bigtable#1#2{\global\npageno=0\advance\tabno by 1
    \setbox\tabbox=\vbox{\tabskip=0pt
    \offinterlineskip\halign{#2}}
       $$\vbox{\ialign{\hfill$##$\crcr
                  \vtop{\hsize=\wd\tabbox\hangindent2truecm\noindent%
                  \hbox to 2truecm{Table \the\tabno\ -- \hfill}#1}\crcr
                  \noalign{\vskip9.5pt}
                  \box\tabbox\crcr}}$$}
\newbox\figbox
\newcount\hgtno
\newcount\basno
\def\figure#1#2#3#4#5{\par\global\advance\figno by 1 \ifnum\prevgraf=0
    \noindent
    \else\newpar\fi \figindent=#1\advance\figindent by 5mm%
    \setbox\figbox=\vbox{\tenpoint\baselineskip11.5pt\hsize
        =#1\hangindent1.75truecm\noindent\hbox to 1.75truecm{Figure \the\figno
        \ -- \hfill}#5}%
    \figheight=#2\advance\figheight by 16pt\advance\figheight by \ht\figbox
    \hgtno=\figheight \basno=\baselineskip \divide\hgtno by \basno
    \advance\hgtno by 1 \advance\hgtno by #3 \advance\figheight by #4%
    \hangindent=\the\figindent\hangafter=-\the\hgtno\noindent
    \hskip-\the\figindent$\smash{\hbox to \the\figindent{\vtop to
    \the\figheight{\hsize=#1\vfill\vskip8pt\vbox{\tenpoint
    \baselineskip11.5pt\hsize=#1\hangindent1.75truecm\noindent
    \hbox to 1.75truecm{Figure \the\figno\ --
    \hfill}#5}\vskip8pt\ }\hfill}}$}
\def\smallboxfigure#1#2#3#4#5{\par\global\advance\figno by 1
\ifnum\prevgraf=0
    \noindent
    \else\smallnewpar\fi \figindent=#1\advance\figindent by 5mm%
    \figwidth=#1\advance\figwidth by -2.4pt
    \setbox\figbox=\vbox{\ninepoint\baselineskip10.35pt\rm\hsize
     =#1\hangindent1.65truecm\noindent\hbox to 1.65truecm{Figure \the\figno
        \ -- \hfill}#5}%
    \figheight=#2\advance\figheight by 16pt\advance\figheight by \ht\figbox
    \hgtno=\figheight \basno=\baselineskip \divide\hgtno by \basno
    \advance\hgtno by 1 \advance\hgtno by #3 \advance\figheight by #4%
    \hangindent=\the\figindent\hangafter=-\the\hgtno
    \newfigheight=#2\advance\newfigheight by -2.4pt
    \noindent \hskip-\the\figindent$\smash{\hbox to \the\figindent{\vtop to
    \the\figheight{\vfill\hsize=#1\vskip-3pt\ialign to
    #1{\vrule##width1.2pt&\hfill\hbox to
    \figwidth{\hfil##\hfil}\hfill&\vrule##width1.2pt\crcr
    \noalign{\hrule height1.2pt}%
    height\newfigheight&\ &height\newfigheight\crcr
    \noalign{\hrule height1.2pt}}\vfill\vskip8pt\vbox{
   \ninepoint\baselineskip10.35pt\rm\hsize=#1\hangindent1.65truecm\noindent
    \hbox to 1.65truecm{Figure \the\figno\ --\hfill}#5}\vskip0pt\
    }\hfill}}$}
\def\nocapsmallfigure#1#2#3#4{\par\global\advance\figno by 1 \ifnum\prevgraf=0
    \noindent
    \else\smallnewpar\fi \figindent=#1\advance\figindent by 5mm%
    \figheight=#2\advance\figheight by 8pt
    \hgtno=\figheight \basno=\baselineskip \divide\hgtno by \basno
    \advance\hgtno by 1 \advance\hgtno by #3 \advance\figheight by #4%
    \hangindent=\the\figindent\hangafter=-\the\hgtno\noindent
    \hskip-\the\figindent$\smash{\hbox to \the\figindent{\vtop to
    \the\figheight{\hsize=#1\vfill}\hfill}}$}
\def\fullfigure#1#2#3#4{\par\global\npageno=0
    \global\advance\figno by 1 \ifnum\prevgraf=0 \noindent
    \else\newpar\fi
    \setbox\figbox=\vbox{\tenpoint\baselineskip11.5pt\hsize
    =#1\hangindent1.75truecm\noindent\hbox to 1.75truecm{Figure \the\figno
        \ -- \hfill}#3}%
    \figheight=#2\advance\figheight by 16pt\advance\figheight by \ht\figbox
    $$\vbox to \figheight{\hsize=#1\vfill#4\vskip8pt\vbox{
     \tenpoint\baselineskip11.5pt\hsize=#1\hangindent1.75truecm\noindent
     \hbox to 1.75truecm{Figure \the\figno\ --\hfill}#3}\vskip2.5pt}$$}
\def\plaktable#1#2#3{\par\global\npageno=0
    \global\advance\tabno by 1 \ifnum\prevgraf=0 \noindent
    \else\newpar\fi
    \setbox\figbox=\vbox{\tenpoint\baselineskip11.5pt\hsize
    =#1\hangindent1.75truecm\noindent\hbox to 1.75truecm{Table
    \the\tabno\ -- \hfill}#3}%
    \figheight=#2\advance\figheight by 16pt\advance\figheight by \ht\figbox
    $$\vbox to \figheight{\hsize=#1\vbox{\tenpoint\baselineskip11.5pt
     \hsize=#1\hangindent1.75truecm\noindent
     \hbox to 1.75truecm{Table \the\tabno\ --\hfill}#3}\vfill}$$}
\def\fullboxfigure#1#2#3{\par\global\npageno=0
    \global\advance\figno by 1 \ifnum\prevgraf=0 \noindent
    \else\newpar\fi
\figwidth=#1\advance\figwidth by -2.4pt
    \setbox\figbox=\vbox{\tenpoint\baselineskip11.5pt\hsize
    =#1\hangindent1.75truecm\noindent\hbox to 1.75truecm{Figure \the\figno
        \ -- \hfill}#3}%
    \figheight=#2\advance\figheight by 24pt\advance\figheight by
\ht\figbox
\newfigheight=#2\advance\newfigheight by -2.4pt
    $$\vbox to
\figheight{\vfill\vskip8pt\ialign to #1{\vrule##width1.2pt&\hfill\hbox to \figwidth{\hfil##\hfil}\hfill&\vrule##width1.2pt\crcr
\noalign{\hrule height1.2pt}
height\newfigheight&\ &height\newfigheight\crcr
\noalign{\hrule height1.2pt}}\vfill\vskip8pt\vbox{
     \tenpoint\baselineskip11.5pt\hsize=#1\hangindent1.75truecm\noindent
     \hbox to 1.75truecm{Figure \the\figno\
     --\hfill}#3}\vskip2.5pt}$$}
\def\smallfullboxfigure#1#2#3{\par\global\npageno=0
    \global\advance\figno by 1 \ifnum\prevgraf=0 \noindent
    \else\newpar\fi
\figwidth=#1\advance\figwidth by -2.4pt
    \setbox\figbox=\vbox{\ninepoint\baselineskip10.35pt\hsize
    =#1\hangindent1.65truecm\noindent\hbox to 1.65truecm{Figure \the\figno
        \ -- \hfill}#3}%
    \figheight=#2\advance\figheight by 24pt\advance\figheight by
\ht\figbox
\newfigheight=#2\advance\newfigheight by -2.4pt
    $$\vbox to
\figheight{\vfill\vskip8pt\ialign to
#1{\vrule##width1.2pt&\hfill\hbox to
\figwidth{\hfil##\hfil}\hfill&\vrule##width1.2pt\crcr
\noalign{\hrule height1.2pt}
height\newfigheight&\ &height\newfigheight\crcr
\noalign{\hrule height1.2pt}}\vfill\vskip8pt\vbox{
     \ninepoint\baselineskip10.35pt\hsize=#1\hangindent1.65truecm\noindent
     \hbox to 1.65truecm{Figure \the\figno\ --\hfill}#3}\vskip2.5pt}$$}
\def\smallfullfigure#1#2#3{\par\global\npageno=0
    \global\advance\figno by 1 \ifnum\prevgraf=0 \noindent
    \else\newpar\fi
    \setbox\figbox=\vbox{\ninepoint\baselineskip10.35pt\hsize
     =#1\hangindent1.65truecm\noindent\hbox to 1.65truecm{Figure \the\figno
        \ -- \hfill}#3}%
    \figheight=#2\advance\figheight by 16pt\advance\figheight by \ht\figbox
    $$\vbox to \figheight{\hsize=#1\vfill\vskip8pt\vbox{
     \ninepoint\baselineskip10.35pt\hsize=#1\hangindent1.65truecm\noindent
     \hbox to 1.65truecm{Figure \the\figno\ --\hfill}#3}\vskip2.5pt}$$}
\def\nocapsmallfullfigure#1#2#3{\par\global\npageno=0
    \global\advance\figno by 1 \ifnum\prevgraf=0 \noindent
    \else\newpar\fi
    \setbox\figbox=\vbox{\ninepoint\baselineskip10.35pt\hsize
     =#1{#3}}%
    \figheight=#2\advance\figheight by 16pt\advance\figheight by \ht\figbox
    $$\vbox to \figheight{\hsize=#1\vfill\vskip8pt\vbox{
     \ninepoint\baselineskip10.35pt\hsize=#1{\noindent #3}}\vskip2.5pt}$$}

\newskip\basskip
\basskip=11.5pt
\def\racco#1{\hbox to 10pt{$\smash{\vcenter{\hbox{$\left.\vphantom{\vcenter{
\vrule height #1\basskip}}\right\}$}}\hfill}$}}
\def\raccol#1{$\smash{\raise 0.5\basskip\hbox{#1}}$}

\def\0{\phantom{0}}
\def\1{\phantom{.0}}
\def\E{\hbox to 6truemm{\hfill E\hfill}}
\def\W{\hbox to 6truemm{\hfill W\hfill}}
\def\N{\hbox to 6truemm{\hfill N\hfill}}
\def\S{\hbox to 6truemm{\hfill S\hfill}}
\mathchardef\g="020E
\mathchardef\h="0068
\mathchardef\m="006D
\mathchardef\s="0073
\newbox\help
\newbox\punt
\newskip\terug
\newskip\vooruit
\def\dg{{}^\g\setbox\help=\hbox{${}^\g$}\setbox\punt=\hbox{$.$}\skip
\terug=-.5\wd\help plus0pt minus0pt\advance\skip\terug by -0.17em plus0em
minus0em\hskip\skip\terug\skip\vooruit=-\skip\terug\advance\skip\vooruit by
-\wd\punt.\hskip\skip\vooruit}
\def\dh{{}^\h\setbox\help=\hbox{${}^\h$}\setbox\punt=\hbox{$.$}\skip
\terug=-.5\wd\help plus0pt minus0pt\advance\skip\terug by -0.17em plus0em
minus0em\hskip\skip\terug\skip\vooruit=-\skip\terug\advance\skip\vooruit by
-\wd\punt.\hskip\skip\vooruit}
\def\dm{{}^\m\setbox\help=\hbox{${}^\m$}\setbox\punt=\hbox{$.$}\skip
\terug=-.5\wd\help plus0pt minus0pt\advance\skip\terug by -0.17em plus0em
minus0em\hskip\skip\terug\skip\vooruit=-\skip\terug\advance\skip\vooruit by
-\wd\punt.\hskip\skip\vooruit}
\def\ds{{}^\s\setbox\help=\hbox{${}^\s$}\setbox\punt=\hbox{$.$}\skip
\terug=-.5\wd\help plus0pt minus0pt\advance\skip\terug by -0.17em plus0em
minus0em\hskip\skip\terug\skip\vooruit=-\skip\terug\advance\skip\vooruit by
-\wd\punt.\hskip\skip\vooruit}
\def\dpr{{}^\prime\setbox\help=\hbox{${}^\prime$}\setbox\punt=\hbox{$.$}\skip
\terug=-.5\wd\help plus0pt minus0pt\advance\skip\terug by -0.17em plus0em
minus0em\hskip\skip\terug\skip\vooruit=-\skip\terug\advance\skip\vooruit by
-\wd\punt.\hskip\skip\vooruit}
\def\ddp{{}^{\prime\prime}\setbox\help=\hbox{${}^{\prime\prime}$}
\setbox\punt=\hbox{$.$}\skip
\terug=-.5\wd\help plus0pt minus0pt\advance\skip\terug by -0.17em plus0em
minus0em\hskip\skip\terug\skip\vooruit=-\skip\terug\advance\skip\vooruit by
-\wd\punt.\hskip\skip\vooruit}
\def\?{\hbox{\rm ?}}%
\def\advancepageno{\global\npageno=1\global\advance\pageno by 1\ifodd\pageno
    \global\hoffset=-1truecm\else\global\hoffset=0truecm\fi}
\ifodd\pageno\global\hoffset=0truecm\else\global\hoffset=-1truecm\fi
\def\makeheadline{\vbox to 0pt{\vskip-1.3truecm
     \line{\vbox to 9.75pt{}\the\headline}\vss}\nointerlineskip}
\newcount\volno
\newcount\nrno
\newcount\yearno
\headline={\elfpoint\sl\ifodd\pageno WGN, the Journal of the IMO
\the\volno:\the\nrno\ (\the\yearno)\hfill$\the\pageno$\else
$\the\pageno$\hfill WGN, the Journal of the IMO
\the\volno:\the\nrno\ (\the\yearno)\fi}
\newcount\footno
\global\footno=0
\def\nfoot#1{\global\npageno=0\advance\footno by 1
    {\tenpoint\baselineskip11.5pt\footnote{$^{\the\footno}$}{#1}}}
\def\smallnfoot#1{\global\npageno=0\advance\footno by 1%
{\ninepoint\baselineskip10.35pt\footnote{$^\the\footno$}{#1}}}%
\def\boxit#1{\vtop{\hrule height1.2pt \hbox{\vrule width1.2pt\kern3.5pt
    \vbox{\kern3.5pt#1\kern3.5pt}
    \kern3.5pt\vrule width1.2pt}\hrule height1.2pt}}
\hsize=17.2truecm
\hfuzz=2pt
\vfuzz=2pt
\vsize=25.2truecm
\voffset=-0.5truecm
\newcount\firstpageno
\newcount\lastpageno
\firstpageno=\pageno
\twelvepoint
\baselineskip 13.75pt
\nopagenumbers
%
\title{The Observation of Lunar Impacts II.}
{Costantino Sigismondi and Giovanni Imponente}
{The frequency and the characteristics of lunar impacting meteorites are reconsidered under the general assumption of belonging to the sporadic meteoroids. We develop the model for evaluating the luminous energy detected in the visual band during the impact. The values obtained are consistent with the luminosity of an Earth's meteor as seen at the Moon's distance, although we recover significantly smaller magnitudes for the lunar impacts with respect to other authors.}

{\tenpoint\rm\baselineskip11.5pt\footnote{}{\hskip-\parindent 
The authors are affiliated with the Department of Physics,
University of Rome ``La Sapienza,'' and ICRA, International Center for
Relativistic Astrophysics, P$^{\rm le}$ A.~Moro 2, I-00185 Rome,
Italy. Costantino Sigismondi can be contacted at {\tt sigismondi@icra.it}.}}%
\section{The detection of lunar impacts}

Although the lunar transient phenomena (LTP) have been observed since several tens of years, it is only recently that they reached the dignity of a scientific problem. Thanks to the effort of some groups of scientists orchestrated by D. Dunham, it was possible to detect unambigously five flashes onto the night side of the Moon [1] during the Leonids meteor shower of 1999.

The opportunity to detect other impacts out of known active showers has been taken into account in our first paper [2].
Our goal there was to evaluate the possibility to have really observed one of them during the total eclipse of the Moon of January $21^{st}$ 2000. A similar approach has been followed by Ortiz et al. [3] in order to explain another possible lunar impact observed on July $16^{th}$ 1999. 

The technological possibility to monitor quite continously the Moon down to magnitudes fainter than the visible limit offered by the CCD videocameras as well as their detection's quantum efficiency larger than the classical photo plates have allowed to attain a rather big number of detected events during the last 12 months. The publishing output at the same period was comparable to the whole activity till then [8].

In this way, one can say that the Moon becomes the best laboratory for studying the meteor showers thanks to its large collecting area with respect to that one coverable by a single observer or even by a network of observers and of instruments devoted to that purpose.

Moreover, during a meteor shower, in some conditions the Moon is visible in the horizon of different places were the zenithal angular distance of the radiant is different. This fact allows to perform a study of the large meteoroid population without considering the problems of the effect of the radiant position in the sky nor the collecting volume effect due to the brightness of the fireballs that are visible at great distances and at low elevations. The latter case occurred  in the fireballs' peak of 1998 Leonids when it was frequently said that it was easier to look toward the horizon for looking more fireballs [4].

Finally the Moon is sampling a region $\sim 400000$ Km apart from the Earth and can intercept the stream in denser regions as happened in 1999 Leonids. A wider knowledge of the entire structure of the stream can be achieved studying the streams both groundbased and looking the dark side of the Moon. Moreover, studies of the temporary sodium atmosphere of the Moon during meteor showers [5] can be joined to the studies of the meteor streams.

\section{The relation between kinetic energy and magnitude of a lunar impact}%
Developing the approach of our first paper [1] we consider 
the formula giving the amount of radiation, assuming the kinetic energy transforms entirely into radiant energy (luminous efficiency $\eta\sim 1$):
$$W_M= {\sigma T^4 \times A_M\over 4\pi\,d_{\rm Moon}^2}, $$
where $A_M \approx (M/\rho)^{2/3}$ is the area of the incoming
meteoroid and $d_{\rm
Moon}\approx 3.84 \times 10^8$~m is the Earth-Moon average distance.
\newpar

In our previous work we assumed an impacting velocity of $v=41$~km/s, obtained averaging the geocentric velocities of all known meteors showers.
Here we extend to the whole spectrum of velocities, and we recover the behaviour of the equation for different values of the mass. Moreover we take into account that the typical velocity for sporadic meteoroids is $\sim 20\div 30$ Km/s [3, 6, 7].

To calculate the visual magnitude, we must take into account that the
eye is sensitive in a range of wavelengths between 400 nm-700nm, with a mean of 550nm. It implies its maximum detection efficiency for a temperature of about 5300 K. 

The kinetic energy in calories (neglecting the melting heats and assuming the calorimetric equation for liquid water in all the range of impacting energies) corresponds to an increment of temperature for each gram of matter equal to
 $\Delta T = v^2\times 1000/2\times 4.18 \sim T$.
Therefore the temperature depends only on the velocity, here measured in Km/s.

Calculating the value of $W_M$ as a function of velocity $v$ and mass $M$ we obtain typical values of $$W_M \approx 3 \times 10^{-8}\ \hbox{\rm W/$m^2$},$$

for a 10 g icy meteoroid impacting at 41 Km/s and producing a $\Delta T \approx 2\times10^5~K$. The eye can detect only $\sim 1/2000000$ of such flux, due to the ratio $(5300/{2\times10^5})^4$, therefore the energy flux in the visual range is $W_M\approx1.5\times 10^{-14}~W/m^2$, i.e., a magnitude 
$$m=-2.5\,\log{1.5\times 10^{-14}\over 3.7\times 10^{-9}}=13.5,$$
where $3.7\times 10^{-9}~W/m^2$ is the visual energy flux
corresponding to a magnitude~0 event. 
That value is consistent with the calculation of the magnitude $m_E$ of a similar object impacting in the Earth's atmosphere using the Arlt and Brown formula [4], as seen at the distance $d_{Moon}$ of the Moon:
$$m_E=40-2.5\log(2.732\times 10^{10}M^{0.92}v_G^{3.91}\times( {d_{atm}}/{d_{Moon}} )^2),$$
where  M the meteoroid mass in grams, and $v_G$
its geocentric velocity in km/s, $d_{atm}\sim~100$ Km is the typical quote of Earth's
atmosphere where the meteor's flashes occurs. 
With the previous parameters the magnitude of the lunar flash should be $m_E\sim 13.8$.
In this case we consider only the general concordance between the calculated magnitudes, even if the efficiency $\eta$ of transformation of kinetic energy into radiation is different (and reliably larger) in the latter formula. 

As a conclusion we consider that a 10 g meteor onto the surface of the Moon can be seen only with rather large amateur telescope ($\sim 40$ cm of diameter) as
claimed during Leonids meteor shower by R. Venable [9].

In our simple model of luminous energy release during an impact once fixed the impacting mass we find luminosities significantly smaller than other authors [10]. It is difficult to explain that difference of more than five magnitudes 
with a larger luminous efficiency for the lunar impacts with respect to the Earth's impacts.

In a following paper we will discuss the observation of the "Padua Impact" during the total eclipse of the Moon of January $21^{st}$, already quoted in [2] and its possible confirmation by a CCD image taken by Gary Emerson [11].

\newpar
{\rm Acknowledgemts}%
\smallnewpar
Thanks to Michael Luciuk for his remarks on our first paper.

\appsection{References}%
\noartref{D.W.~Dunham}{{\rm in} IAU Circular\/\rm\ 7320}{1999}%
\noartref{C.~Sigismondi, G.~Imponente}{WGN\/\rm\ 28:2-3}{June 2000, p.~54}%
\noartref{J.~Ortiz, P.~V.~Sada, L.~R.~Bellot Rubio, F.~J. Aceituno, J. Aceituno, P.J. Gutierrez, U. Thiele}
{Nature\/\rm\ 405}{22 June 2000, p.~921}%
\noartref{R.~Arlt, P.~Brown}{WGN\/\rm\ 27:6}{December 1999, p.~278}%
\noartref{S.~Verani, C.~Barbieri, C. Benn, G. Cremonense}{Planet. Space Sci.\/\rm\ 46}{1998, p.~1003}%
\noartref{Z. Ceplecha}{Astronomy and Astrophysics\/\rm\ 286}{1994, p.~967}%
\noartref{P.~Jenniskens}{\rm http://www-space.arc.nasa.gov/\~{}leonid/guide/p1-magnitude.html~}{1998}%
\noartref{D.~Dunham}{\rm http://iota.jhuapl.edu~}{2000}%
\noartref{D.~Dunham}{\rm http://iota.jhuapl.edu/lunar\_~leonid/lunpr501.htm~}{2000}%
\noartref{D.~Dunham}{\rm http://iota.jhuapl.edu/lunar\_~leonid/sizes.htm~}{2000}%
\noartref{D.~Dunham}{\rm http://iota.jhuapl.edu/lunar\_~leonid/emrsn621.htm~}{2000}%
\bye

\end{thebibliography}
\end{document}